\documentclass{jetp-preprint}
\twocolumn

\begin{document}
\English

\title{Specific features of $g\approx4.3$ EPR line behavior in~magnetic nanogranular composites}

\setaffiliation1{P.L.Kapitza Institute for Physical Problems, RAS, Moscow, 119334 Russia}
\setaffiliation2{National Research University Higher School of Economics, Moscow, 101000 Russia}
\setaffiliation3{Voronezh State Technical University, Voronezh, 394026 Russia}
\setaffiliation4{National Research Centre Kurchatov Institute, Moscow, 123182 Russia}
\setaffiliation5{Kotelnikov Institute of Radio Engineering and Electronics, Fryazino Branch, RAS, Fryazino, Moscow oblast, 141190 Russia}

\setauthor{A.~B.}{Drovosekov}{1}\email{drovosekov@kapitza.ras.ru}
\setauthor{N.~M.}{Kreines}{1}
\setauthor{D.~A.}{Ziganurov}{12}
\setauthor{A.~V.}{Sitnikov}{34}
\setauthor{S.~N.}{Nikolaev}{4}
\setauthor{V.~V.}{Rylkov}{45}

\rtitle{Specific features of $g\approx4.3$ EPR line \dots}

\abstract{Films of metal-insulator nanogranular composites M$_x$D$_{100-x}$ with different composition and percentage of metal and dielectric phases (M = Fe, Co, CoFeB; D = Al$_2$O$_3$, SiO$_2$, LiNbO$_3$; $x\approx15{-}70$~at.\,\%) are investigated by magnetic resonance in a wide range of frequencies ($f=7{-}37$~GHz) and temperatures ($T=4.2{-}360$~K). In~addition to the usual ferromagnetic resonance signal from an array of nanogranules, the experimental spectra contain an additional absorption peak, which we associate with the electron paramagnetic resonance (EPR) of~Fe and Co ions dispersed in the insulating space between the granules. In contrast to the traditional EPR of~Fe and Co ions in weakly doped non-magnetic matrices, the observed peak demonstrates a number of unusual properties, which we explain by the presence of magnetic interactions between ions and granules.}

\maketitle

\section{Introduction}

Magnetic nanoparticles and nanogranular systems have long been a subject of intensive research, due to unusual physical properties of these objects, as well as the wide possibilities of applications \cite{Dormann1992,Gubin2009,Bedanta2013}.

Magnetic metal-insulator nanogranular composites (nanocomposites) are an array of ferromagnetic (FM) nanogranules randomly distributed in a solid dielectric medium (matrix). In~our previous works \cite{Rylkov2019, DrovJMMM, DrovJETPlett, Drov2022-1, Drov2022-2}, we~studied nanocomposite films of composition M$_x$D$_{100-x}$ based on the metal FM alloy M = Co$_{40}$Fe$_{40}$B$_{20}$ (hereinafter CoFeB for brevity) and dielectrics D = Al$_2$O$_3$ and LiNbO$_3$. The value $x$ in the structural formula M$_x$D$_{100-x}$ reflects the nominal content of the metal phase in the nanocomposite, a significant part of which forms FM nanogranules. At the same time, a specific feature of the studied films was the high concentration of paramagnetic (PM) ions Fe and Co dispersed in the insulating space between CoFeB granules \cite{RylkovJMMM, RylkovIEEE, Rylkov2020}. It~has been shown that the presence of such ions in the dielectric gaps between the granules leads to the manifestation of unusual electrical, magnetic and magnetoresonance properties of nanocomposites, due to the increase of intergranular electron tunneling and exchange interaction \cite{RylkovPRB, Rylkov2018, Rylkov2019, Rylkov2020, Rylkov2021, GanshinaIEEE, DrovJMMM, DrovJETPlett}.

In the works \cite{Drov2022-1,Drov2022-2} nanocomposite films (CoFeB)$_x$(Al$_2$O$_3$)$_{100-x}$ and (CoFeB)$_x$(LiNbO$_3$)$_{100-x}$ ($x\approx30{-}55$~at.\,\%) were studied by magnetic resonance in a wide range of frequencies ($f=7{-}37$~GHz) and temperatures ($T=4.2{-}360$~K). It was found that in addition to the usual ferromagnetic resonance (FMR) signal from an ensemble of FM granules, the experimental spectra of the films contained an additional much weaker absorption peak. The frequency-field dependence $f(H)$ for this peak proved to be linear in the high frequency region and characterized by an effective $g$-factor $g\approx4.3$ typical of electron paramagnetic resonance (EPR) of Fe$^{3+}$ ions in amorphous solids \cite{Castner1960, Wickman1965, Kliava1988}. It should be noted that such resonance is often observed while studying the iron-based nanoparticles in various media \cite{Koksharov2001, Jitianu2002, Kliava2011, Edelman2012, Ivanova2018}. However, unlike the traditional EPR of Fe$^{3+}$ ions in weakly doped matrices, in our case the observed peak demonstrates a number of unusual properties:
\begin{itemize}
\item its intensity proves to be approximately the same in the transverse and longitudinal geometry of the resonance excitation;
\item the dependence $f(H)$ demonstrates the presence of a finite frequency in the zero field, which increases with the growth of the FM phase content;
\item the position of the peak depends on the orientation of the magnetic field with respect to the film plane;
\item as temperature decreases, the peak shifts towards weaker fields and decreases in intensity, disappearing at $T\lesssim60$~K.
\end{itemize}

Nanogranular films have previously been studied by magnetic resonance in a number of works \cite{Tomita2004, Timopheev2012, Chekrygina2014, Buravtsova2004, Iskhakov2010, Kotov2023, Kotov2020, Lesnik2003, Pires2006, Kakazei2005, Vyzulin2006, Wang1995, Butera1999, Gomez2004, Denisova2011, Denisova2012, Kablov2016, Denisova2017}. Often, besides the main FMR line, the authors observed additional absorption peaks, which were associated either with the inhomogeneity of the samples \cite{Kotov2020, Lesnik2003, Pires2006}, or~with the excitation of inhomogeneous oscillations in the films \cite{Kakazei2005, Vyzulin2006}, in particular spin-wave and surface modes \cite{Wang1995, Butera1999, Gomez2004, Denisova2011, Denisova2012, Kablov2016, Denisova2017}. Observation of several resonance peaks caused by the excitation of inhomogeneous modes is also possible in ordered arrays of magnetic nanoparticles \cite{Martyanov2005, Martyanov2015, Kakazei2015, Neugebauer2020}.

In our case, the behavior of the additional peak does not agree with the described scenarios. It proves to be more productive to assume that this peak is associated with EPR of Fe$^{3+}$ ($g\approx4.3$) ions dispersed in the insulating space between FM granules. As shown in \cite{Drov2022-1}, the frequency and orientational dependencies of the resonance field for the additional peak are well described taking into account the shift of the EPR frequency due to dipole-dipole and exchange interactions of the Fe$^{3+}$ ions with the ensemble of FM granules. However, the proposed model does not entirely explain the non-standard excitation conditions of this peak and its anomalous behavior with temperature \cite{Drov2022-2}.

In order to further clarify the reasons for the anomalous behavior of the EPR peak, in this work, we study a wider set of nanocomposites with various compositions M$_x$D$_{100-x}$. Besides the systems based on the CoFeB alloy, we investigate films in which FM granules are formed of pure iron or cobalt in different insulating matrices D = Al$_2$O$_3$, SiO$_2$, LiNbO$_3$. Furthermore, the range of the FM phase content ($x \approx 15{-}70$~at.\,\%) is significantly expanded as compared with previous works.

\section{Samples, their preliminary characterization, methods}

Nanocomposite films M$_x$D$_{100-x}$ with a thickness of $\approx1{-}3$~$\mu$m were synthesized by ion beam sputtering on glass-ceramic substrates using composite targets \cite{Granovsky2016, Stognei2014, Kalinin2007}. The target represents a plate of FM metal Fe, Co or an alloy Co$_{40}$Fe$_{40}$B$_{20}$ (CoFeB), on which a number of rectangular strips of oxides Al$_2$O$_3$, SiO$_2$ or LiNbO$_3$ are placed. The uneven arrangement of dielectric strips on the target surface allows the formation of a nanocomposite film M$_x$D$_{100-x}$ with a smooth controlled change in the concentration $x$ along the substrate in a wide range $\Delta x\approx 30{-}40$~at.\,\%. Further studies are carried out on individual pieces of the grown film with a size of $5\times5$~mm$^2$, so that the change of $x$ within one sample is less than 1~at.\,\%. The content of the metal phase in the films was determined by energy dispersive X-ray microanalysis. The following series of samples were studied:

(CoFeB)$_x$(Al$_2$O$_3$)$_{100-x}$, \hfill $x \approx 15{-}56$~at.\,\%; \hskip 8mm ~

Fe$_x$(Al$_2$O$_3$)$_{100-x}$, \hfill $x \approx 31{-}58$~at.\,\%; \hskip 8mm ~

(CoFeB)$_x$(SiO$_2$)$_{100-x}$, \hfill $x \approx 20{-}67$~at.\,\%; \hskip 8mm ~

Co$_x$(SiO$_2$)$_{100-x}$, \hfill $x \approx 24{-}67$~at.\,\%; \hskip 8mm ~

(CoFeB)$_x$(LiNbO$_3$)$_{100-x}$, \hfill $x \approx 30{-}48$~at.\,\%; \hskip 8mm ~

Co$_x$(LiNbO$_3$)$_{100-x}$, \hfill $x \approx 33{-}41$~at.\,\%; \hskip 8mm ~

According to transmission electron microscopy data, the obtained composites consist of crystalline FM nanogranules randomly distributed inside the amorphous oxide matrix \cite{RylkovPRB, Kalinin2007, Tregubov2013, Denisova2018}. The granules usually have approximately spherical shape with diameter $\approx2{-}8$~nm depending on the composition and concentration $x$. However, in the case of the LiNbO$_3$ matrix, the granules tend to stretch in the direction of film growth up to $\approx20$~nm \cite{Rylkov2018, Rylkov2021}. The structures Co$_x$(LiNbO$_3$)$_{100-x}$ are characterized by strongly inhomogeneous distribution of the shape and size of granules over the film thickness \cite{Rylkov2021}.

Depending on the composition, the percolation threshold $x_p$ of nanocomposites differs within the range $x_p\approx45{-}60$~at.\,\% \cite{RylkovPRB, Rylkov2018, Rylkov2021, Iskhakov2010, Stognei2014, Kalinin2007}. Slightly below this threshold at $x_c<x<x_p$ ($x_p-x_c\approx5{-}10$~at.\,\%), the films demonstrate an interesting logarithmic temperature dependence of the conductivity \cite{Rylkov2020, RylkovPRB, Rylkov2018, Rylkov2021}, typical of granular systems with ``strong tunnel coupling'' between the granules \cite{Efetov2003}. According to magnetic data, approximately in this concentration range, the samples demonstrate a transition from superparamagnetic to ferromagnetic behavior \cite{DrovJMMM, GanshinaIEEE, Chekrygina2014, Buravtsova2004, Iskhakov2010, Kotov2023, Denisova2011}. At the same time, a sharp increase in the magnetization of the films is observed in the low temperature region, which is explained by a large number of magnetic ions Fe and Co dispersed in the insulating matrix \cite{RylkovJMMM, RylkovIEEE, Rylkov2020, RylkovPRB}.

In this work, the nanocomposite samples are studied by magnetic resonance in a wide range of frequencies ($f=7{-}37$~GHz) and temperatures ($T=4.2{-}360$~K) using a laboratory transmission-type spectrometer based on rectangular and tunable cylindrical resonators \cite{Drov2022-1}. The experiments are carried out at different orientations of the external field $\mathbf{H}$ (up to 17~kOe) with respect to the film plane. Note that in~the case of in-plane field, it is possible to realize both transverse ($\mathbf{h}\perp\mathbf{H}$) and longitudinal ($\mathbf{h}\parallel\mathbf{H}$) geometry of resonance excitation by the microwave field $\mathbf{h}$ \cite{Drov2022-1}.

\section{Magnetic resonance spectra and~their discussion}

\subsection{The case of in-plane field (room~temperature)}

Almost for all samples, regardless of composition, the qualitative behavior of the magnetic resonance spectra looks identical (Fig.~1). In usual transverse resonance excitation geometry ($\mathbf{h}\perp\mathbf{H}$), one intensive FMR peak is observed as a rule. The width of this peak varies for different compositions. In the case when the FMR peak is sufficiently narrow, as for the system (CoFeB)$_x$(Al$_2$O$_3$)$_{100-x}$, in lower fields it is possible to resolve a second, less intensive absorption peak, which we associate with EPR of magnetic ions dispersed in the insulating matrix \cite{Drov2022-1}. When the resonance excitation geometry changes to a longitudinal one ($\mathbf{h}\parallel\mathbf{H}$), the intensity of the FMR peak drops significantly, which is natural, but the amplitude of the EPR peak remains approximately the same. As~a~result, the EPR peak is much better manifested in the geometry $\mathbf{h}\parallel\mathbf{H}$ and is reproduced almost for all compositions of the films, with the exception of Co$_x$(LiNbO$_3$)$_{100-x}$ \cite{Drov2022-1}.

\begin{figure}
\centering
\includegraphics[width=0.9\columnwidth]{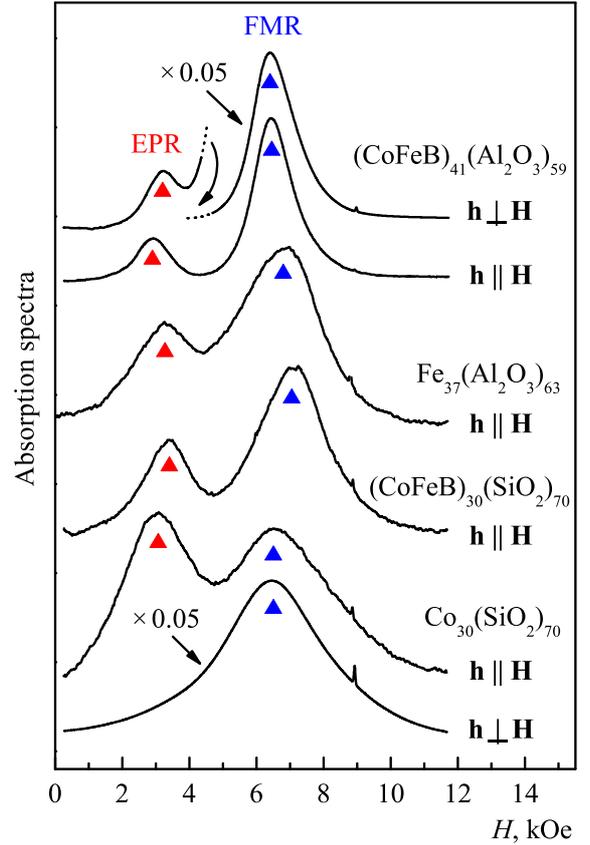}
\caption{Room temperature spectra for nanocomposite films M$_x$D$_{100-x}$ ($x \approx 30{-}40$\,at.\%) with different compositions of metal and dielectric phase (M and D). Spectra are obtained in magnetic field applied in the film plane at~frequency $f\approx25$~GHz in transverse ($\mathbf{h}\perp\mathbf{H}$) and longitudinal ($\mathbf{h}\parallel\mathbf{H}$) geometries of resonance excitation.}
\end{figure}

\begin{figure}
\centering
\includegraphics[width=0.9\columnwidth]{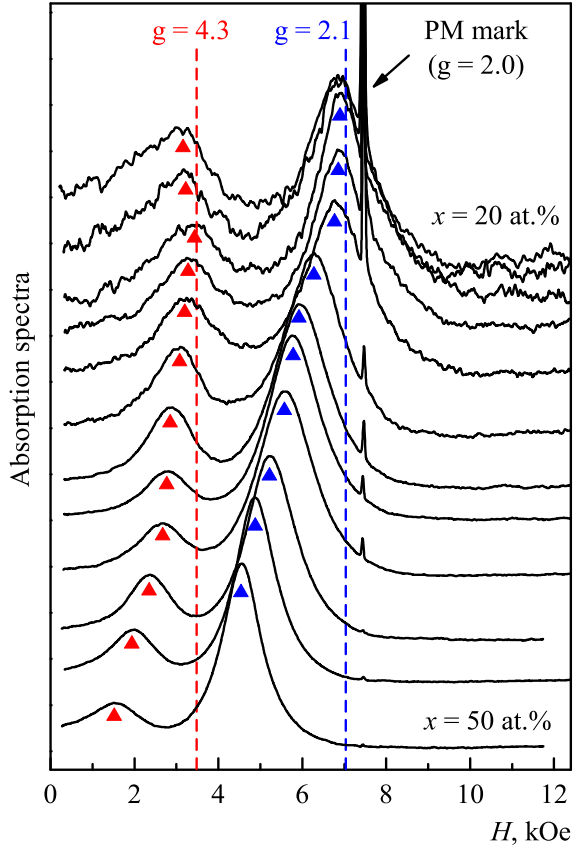}
\caption{Spectra for films (CoFeB)$_x$(Al$_2$O$_3$)$_{100-x}$ with different content of the FM phase $x$. Spectra are obtained in magnetic field applied in the film plane in longitudinal geometry of resonance excitation ($\mathbf{h}\parallel\mathbf{H}$) at frequency $f\approx21$~GHz ($T=296$~K). Spectra are normalized on the FMR peak amplitude.}
\end{figure}

\begin{figure}
\centering
\includegraphics[width=0.9\columnwidth]{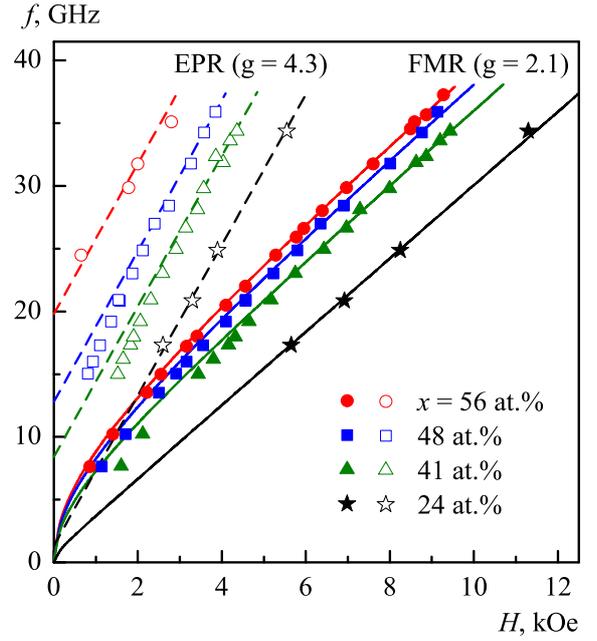}
\caption{Frequency-field dependencies $f(H)$ for FMR and EPR peaks in films (CoFeB)$_x$(Al$_2$O$_3$)$_{100-x}$ with different content of the FM phase $x$. Magnetic field is applied in the film plane ($T=296$~K). Points are the experimental data, solid lines correspond to Kittel formula (1), dashed lines are linear dependencies (2).}
\end{figure}

The position and intensity of the EPR peak depends on the content of the metal FM phase $x$ in the films. Interestingly, it is best manifested at concentrations well below the percolation threshold $x_p$. When approaching this threshold, the intensity of the EPR peak decreases, and completely vanishes beyond $x_p$.

The detailed concentration dependence of the magnetic resonance spectra was obtained for the system (CoFeB)$_x$(Al$_2$O$_3$)$_{100-x}$ (Fig.~2). In the limit of low concentrations $x<25$~at.\,\% the FMR line approaches the nominal position corresponding to $g$-factor $g\approx2.1$, which is characteristic of bulk FM metals Fe, Co and their alloys. This peak is obviously associated with the resonance from the array of weakly interacting FM nanoparticles. Note that instead of the term ``FMR'', the term ``superparamagnetic resonance'' can be used in this case \cite{Kliava2011}.

The position of the second peak at $x<25$~at.\,\% corresponds to the effective $g$-factor $g\approx4.3$, characteristic of EPR for Fe$^{3+}$ ions in amorphous solids. Note that the obtained type of magnetic resonance spectra for nanocomposites in the limit of low FM phase concentrations is quite consistent with the results of other authors for iron-based nanoparticles in various media \cite{Koksharov2001, Jitianu2002, Kliava2011, Edelman2012, Ivanova2018}. In our case, however, there is an interesting trend that the EPR peak with $g\approx4.3$ is better manifested in the unusual longitudinal geometry of resonance excitation ($\mathbf{h}\parallel\mathbf{H}$).

Note another not quite expected result, that in our case the EPR peak ($g\approx4.3$) is observed not only for systems based on iron-containing granules Fe and CoFeB, but also for the cobalt-based nanocomposite Co$_x$(SiO$_2$)$_{100-x}$ (Fig.~1). The theory of EPR predicts the possibility of effective $g$-factor $g\approx4.3$ for Co$^{2+}$ ions in the case of octahedral ligand field \cite{Abragam1972}. Experimentally, such a line is also sometimes observed in some cubic crystals or nanocrystallites with Co ions \cite{Legein1993, Raita2011, Mesaros2014}. Perhaps a similar scenario is realized in our case. It~is also possible that the observed lines with $g\approx4.3$ are caused by the excitation of ``forbidden'' transitions between spin states of PM centers with the change of the spin projection $\Delta m_S=\pm2$. Note that transitions of this type can be excited by both transverse and longitudinal microwave magnetic field~\cite{Abragam1972}.

An increase of the FM phase content in the films leads to the growth of magnetodipole interactions in the system and the appearance of significant demagnetizing fields. In this situation, the FMR line shifts towards weaker fields (Fig.~2). At the same time, a~similar shift of the resonance field is also observed for the EPR peak.

Figure~3 shows the frequency-field dependencies $f(H)$ for both absorption peaks in films (CoFeB)$_x$(Al$_2$O$_3$)$_{100-x}$. At low FM phase content $x<25$~at.\,\% these dependencies are close to linear with effective $g$-factors $g\approx2.1$ and $g\approx4.3$ for FMR and EPR peaks, respectively. At higher concentrations~$x$, the dependence $f(H)$ for the FMR peak is described by the Kittel formula
\begin{equation}
f = \gamma_\mathrm{FMR} \sqrt{H(H+4\pi M)},
\end{equation}
where the gyromagnetic ratio $\gamma_\mathrm{FMR}\approx2.92$~GHz/kOe corresponds to $g$-factor $g\approx2.1$, and the effective demagnetizing field of the film $4\pi M$ increases with the growth of $x$.

For the EPR peak, the dependence $f(H)$ in the high frequency region is described by a linear function
\begin{equation}
f = \gamma (H + \delta H),
\end{equation}
where the gyromagnetic ratio $\gamma\approx 6.0$~GHz/kOe corresponds to $g$-factor $g\approx4.3$, and the line shift $\delta H$ increases with the growth of $x$.

According to a simple model proposed in \cite{Drov2022-1}, the line shift $\delta H$ in formula (2) arises due to the interaction of PM ions with FM granules and is determined by the effective field
\begin{equation}
\delta H = J M,
\end{equation}
which acts on the PM ion from the ensemble of FM granules ($M$ is average magnetization of the array of FM granules, $J$ is the effective field constant).

In \cite{Drov2022-1,Drov2022-2} this field was associated with the exchange interaction between ions and granules. In this case, the dimensionless constant $J$ has the meaning of the effective exchange field parameter. As will be shown below, an alternative explanation of the effective field $JM$ of a dipole-dipole nature is possible.

\subsection{The diagram $\delta H - 4\pi M$. Manifestation~of~the~``Lorentz field''?}

Equations (1)$-$(3) qualitatively explain the correlated shift of FMR and EPR lines to low fields with an increase of the concentration $x$. Indeed, in both cases, the shift of the absorption peak is determined by the average magnetization of the film $M$. The values $4\pi M$ and $\delta H$ can be determined experimentally from the positions of the FMR and EPR peaks, respectively. In this case, the effective constant $J$ is determined by the ratio $J/4\pi = \delta H / 4\pi M$.

It should be noted that the effective field $4\pi M$ in formula (1), generally speaking, may depend on the shape and anisotropy of the FM granules and differ from the static value $4\pi M$ of the film. The equivalence of these values can be expected in the case of spherical granules in the absence of any preferred anisotropy axis \cite{Dubowik1996, Ignatchenko2010}. In our case, this condition seems to be fulfilled for most structures, with the exception of nanocomposites based on the LiNbO$_3$ insulating matrix, where the granules have a shape elongated in the direction of film growth \cite{Rylkov2019, RylkovJMMM, Rylkov2021}.

Assuming the exchange nature of the effective field $\delta H$, one would expect that the constant $J$ in equation~(3) should depend on many factors: the chemical composition of the granules and the dielectric matrix, the content of the FM phase in the nanocomposite, temperature. However, this constant proves to be quite universal.

Figure~4 shows a summary diagram $\delta H - 4\pi M$ for all the films studied. It can be seen that, regardless of the specific composition of the films, the experimental points lie near some universal linear dependence. This universalism means that the shift of the EPR field $\delta H$ is mainly determined by the average magnetization of the film, and therefore has a dipole-dipole origin, similar to the demagnetization field $4\pi M$. The ratio $\delta H / 4\pi M$ is found to be about 1/3, that is, $\delta H\approx 4\pi M/3$, and the constant $J$ approximately corresponds to the demagnetizing factor of the sphere $J \approx 4\pi/3$.

\begin{figure}
\centering
\includegraphics[width=0.9\columnwidth]{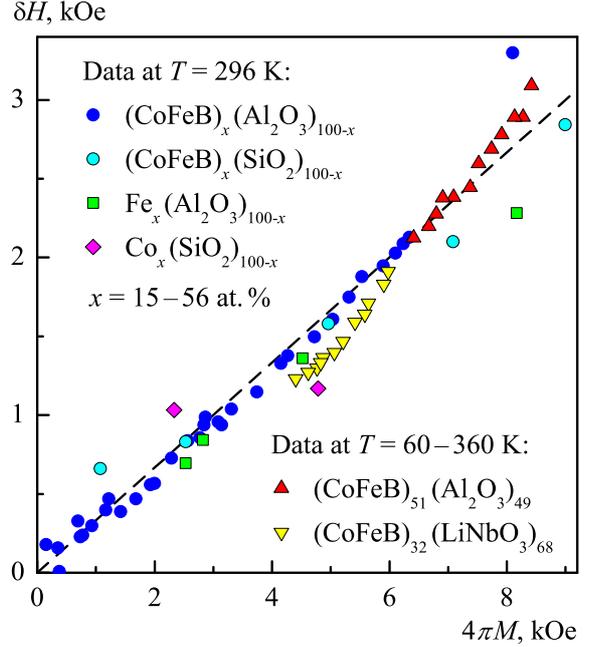}
\caption{Summary diagram $\delta H - 4\pi M$ for all investigated samples. Points are the experimental data obtained in the present work and in previous works \cite{Drov2022-1,Drov2022-2}. Dashed line corresponds to the dependence $\delta H=4\pi M/3$.}
\end{figure}

This result suggests that the EPR peak shift $\delta H$ is associated with the so-called ``Lorentz field'' of dipole nature, which acts on PM ions from the ensemble of FM granules. The concept of the Lorentz field arises in the problem of calculating local fields inside a (quasi-)\,continuous medium, taking into account its real inhomogeneity at the microscopic level. This concept is better known from general physics courses, where it is applied to the calculation of local electric fields in dielectric crystals \cite{Kittel1978, Tamm2003}, as well as to the determination of local magnetic fields in magnetically ordered media by NMR \cite{Turov1969} and muon spectroscopy \cite{Smilga1991}. Its application to the magnetism of nanostructures is sometimes discussed in theoretical works \cite{Dubowik1996, Ignatchenko2010, Bowden2019}. A possible experimental manifestation of the Lorentz field in such systems was considered in \cite{Godinho1995, Dormann1995} while studying the temperature dependence of the susceptibility of nanogranular films.

The Lorentz method of calculating the local field $\mathbf{H}_{loc}$ at some selected point inside an inhomogeneous medium suggests the decomposition of this field into several components:
\begin{equation}
\mathbf{H}_{loc} = \mathbf{H} + \mathbf{H}_{dem} + \mathbf{H}_{L} + \sum \mathbf{H}_{dip}.
\end{equation}
Here $\mathbf{H}$ is the external magnetic field, $\mathbf{H}_{dem}$ is the demagnetization field associated with the shape of the sample. For a thin film, this term has the form
$$
\mathbf{H}_{dem} = -4\pi \mathbf{M}_\perp,
$$
where $\mathbf{M}_\perp$ is the vector component of magnetization normal to the film plane. The third term in (4) $\mathbf{H}_{L}$ is the Lorentz field created by a fictitious spherical cavity (``Lorentz sphere'') cut out in a magnetized medium around the selected point. It is determined by the well-known formula
$$
\mathbf{H}_{L} = \frac{4\pi}{3}\mathbf{M}.
$$
Finally, the last term in (4) is the sum of the dipole fields $\mathbf{H}_{dip}$ created at the selected point by magnetic dipoles (FM granules) located inside the Lorentz sphere.

The EPR frequency of magnetic ions is determined by the local fields on each of them, taking into account these four contributions: $f=\gamma|\mathbf{H}_{loc}|$. Note, that the last term in (4) has random nature due to the disorder in the granules arrangement inside the Lorentz sphere. It can be assumed that the average value of this contribution over all PM centers is zero. This assumption seems to be reasonable in the case of isotropic distribution of spherical granules in the dielectric matrix \cite{Tamm2003, Usov2020}. In this situation, the last term in (4) leads only to a broadening of the resulting EPR line, while the shift of the absorption peak is determined by the first three terms:
\begin{equation}
f = \gamma |\mathbf{H} - 4\pi \mathbf{M}_\perp + 4\pi\mathbf{M}/3|.
\end{equation}
In the case of in-plane orientation of the magnetic field, the demagnetizing field is absent ($\mathbf{M}_\perp=0$). At the same time the situation $\mathbf{H}\parallel\mathbf{M}$ is realized, and equation (5) transforms to
\begin{equation}
f = \gamma (H + 4\pi M/3).
\end{equation}
Thus, the presence of a Lorentz field in a granular medium can explain the shift of the EPR peak $\delta H = 4\pi M/3$. Experimentally observed deviations from this value, which are seen in Fig.~4 as a scatter of points relative to the theoretical straight line, can be caused by various reasons: not quite accurate account for the contribution of the fourth term in equation (4), the non-spherical shape of the granules, the presence of additional exchange interactions.

Note that the best agreement with the model is achieved for nanocomposite films (CoFeB)$_x$(Al$_2$O$_3$)$_{100-x}$, which are characterized by a close to spherical shape of granules with small diameters $2{-}4$~nm \cite{RylkovJMMM, RylkovPRB}. Films of this composition are also distinguished by the narrowest resonance peaks (Fig.~1), indicating a higher degree of homogeneity of the system.

Note that the presence of random magnetic interactions in the system leads to deviations of local fields at PM centers from the direction of the external magnetic field. In this situation, EPR can be excited not only by transverse, but also by longitudinal microwave field, which behavior is observed experimentally.

\subsection{The case of out-of-plane field (room~temperature)}

According to equation (5), in the case of deviation of the magnetic field from the film plane by an arbitrary angle $\theta_H$, it is necessary to take into account the additional contribution to the EPR frequency associated with the appearance of the demagnetizing field $4\pi \mathbf{M}_\perp$. In the particular case of normal orientation of the field, when $\mathbf{H}\parallel\mathbf{M} = \mathbf{M}_\perp$, the frequency-field dependence takes the form
\begin{equation}
f = \gamma (H - 8\pi M/3).
\end{equation}
Thus, comparing with the in-plane geometry, the EPR peak is shifted towards stronger fields. This effect is observed experimentally (Fig.~5). Note that when the magnetic field deviates from the film plane, the FMR line also shifts towards stronger fields. In a normal field, the position of the FMR peak is determined by the well-known Kittel formula:
\begin{equation}
f = \gamma_\mathrm{FMR} (H - 4\pi M).
\end{equation}

As it was shown in \cite{Drov2022-1}, the angular dependence of the EPR field $H_{res}(\theta_H)$ can be calculated analytically, neglecting the dependence of the film magnetization on the magnetic field. Such approximation can be considered adequate for films with sufficiently high content of the FM phase in not too weak magnetic fields, when the effects of superparamagnetism can be disregarded. In the work \cite{Drov2022-1}, a good agreement between the calculated and experimental angular dependencies $H_{res}(\theta_H)$ was demonstrated.

Let us show that taking into account the superparamagnetism of films and the presence of the Lorentz field, the frequency-field dependencies for the EPR peak can be well described in cases of in-plane and normal field, using formulas (6), (7).

Figure 6a shows the experimental dependencies $f(H)$ for FMR and EPR peaks obtained in two geometries for one of the samples (CoFeB)$_x$(Al$_2$O$_3$)$_{100-x}$. The dashed lines correspond to the calculation according to Kittel equations (1), (8) for the FMR peak and (6), (7) for the EPR peak under the assumption of constant value $4\pi M=4\pi M_S$ (``ideal'' FM film). It can be seen that in this case the formulas work only in the region of high frequencies (strong fields). Deviations in weak fields are explained by a decrease of $4\pi M$ as compared to the saturation value $4\pi M_S$.

\begin{figure}
\centering
\includegraphics[width=0.9\columnwidth]{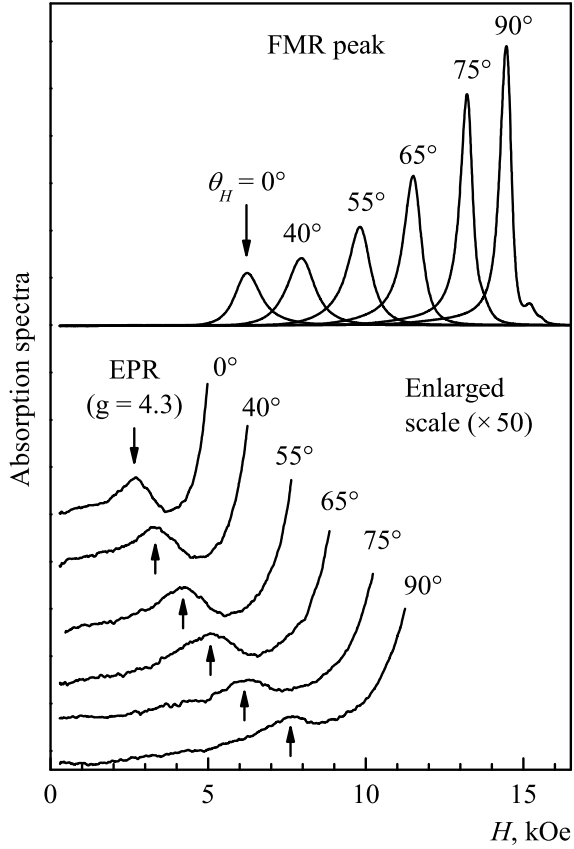}
\caption{Magnetic resonance spectra in nanocomposite (CoFeB)$_{47}$(Al$_2$O$_3$)$_{53}$ for different orientations of the magnetic field with respect to the film plane~($\theta_H$). Spectra are obtained at room temperature at frequency $f\approx25$~GHz in transverse geometry of resonance excitation ($\mathbf{h}\perp\mathbf{H}$).}
\end{figure}

\begin{figure}
\centering
\includegraphics[width=0.9\columnwidth]{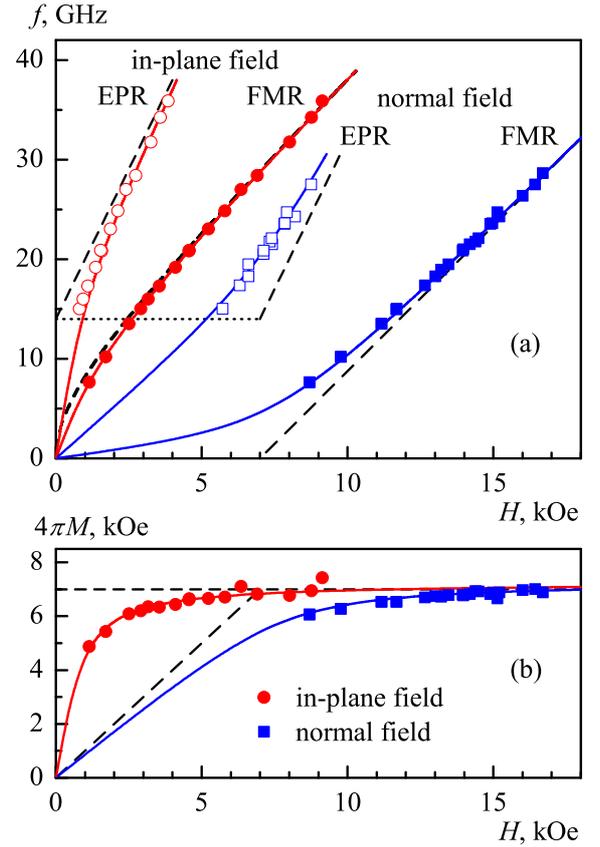}
\caption{(a) Frequency-field dependencies $f(H)$ for FMR and EPR peaks in the film (CoFeB)$_{48}$(Al$_2$O$_3$)$_{52}$ in the cases of in-plane and normal field ($T=296$~K). (b) Field dependencies of $4\pi M$ value. Points are the experimental data, lines are calculations. The dashed lines correspond to the case of ``ideal'' FM film with $4\pi M = 4\pi M_S$, the solid lines are obtained taking into account the filed dependence of $4\pi M$ value (see text).}
\end{figure}

In Fig.~6b, the field dependencies $4\pi M(H)$ for in-plane and normal geometries are recalculated from experimental FMR data using Kittel formulas (1), (8). In~the case of in-plane geometry, the resulting dependence $4\pi M(H)$ can be well described by the usual Langevin function $L(x)$ for superparamagnets
$$
4\pi M = 4\pi M_S \cdot L \left( \frac{\mu H}{k_BT} \right),
$$
where $\mu$ is the magnetic moment of FM granules, $k_B$ is the Boltzmann constant (for the considered nanocomposite $\mu\approx 10^4\mu_B$ Bohr magneton). In normal geometry, the experimental dependence $4\pi M(H)$ can be well approximated by a similar equation with the replacement of $H$ by the ``internal'' field $H_{in}=H-4\pi M$:
$$
4\pi M = 4\pi M_S\cdot L\left( \frac{\mu(H-4\pi M)}{k_BT} \right)
$$
with the same parameters $\mu$ and $4\pi M_S$. In this case, the function $4\pi M(H)$ is not expressed explicitly, but can be defined parametrically considering the value $H_{in}$ as a parameter.

Taking into account the calculated field dependencies of $4\pi M$ in two geometries, the resulting dependencies $f(H)$ obtained by Kittel formulas (1), (8) for the FMR peak and (6), (7) for the EPR peak are plotted  in Fig.~6a by solid lines. It can be seen that taking into account the Lorentz field in equations (6), (7) provides a close to perfect approximation of the experimental behavior for the EPR peak.

\subsection{Temperature evolution of the spectra}

In the work \cite{Drov2022-2}, we investigated the temperature evolution of magnetic resonance spectra in nanocomposite films (CoFeB)$_x$(Al$_2$O$_3$)$_{100-x}$ with FM phase content $x\approx 47{-}51$~at.\,\% and (CoFeB)$_x$(LiNbO$_3$)$_{100-x}$ with $x \approx 30{-}40$~at.\,\%. Qualitatively, the samples showed identical behavior. In the case of magnetic field applied in the film plane, temperature decrease from 360 to 4.2~K initiates a monotonic shift of the FMR peak towards weaker fields, which behavior is explained by an increase of the film magnetization and, as a consequence, the demagnetizing field $4 \pi M$. At the same time, the EPR peak ($g\approx4.3$) also shifts to weaker fields in accordance with the formula $\delta H\approx JM$, where $J\approx 4\pi/3$ (see~Fig.~4).

An unusual feature in the behavior of the EPR peak was that its intensity reduced with decreasing temperature, and below $T\approx60$~K it disappeared. In the present work, it is found that the similar behavior of the EPR line ($g\approx4.3$) is reproduced for films with different compositions M$_x$D$_{100-x}$ in the case of sufficiently high content of the FM phase $x\gtrsim30$~at.\,\%. The observed reduction of the EPR peak intensity with temperature decrease contradicts the typical situation for systems weakly doped with Fe and Co ions, where the weakening of the EPR line occurs, on the contrary, with temperature increase \cite{Koksharov2001, Jitianu2002, Kliava2011, Edelman2012, Legein1993, Raita2011, Mesaros2014}.

It turned out that for the studied nanocomposites, the transition to the limit of low FM phase contents $x\lesssim30$~at.\,\% leads to a more traditional temperature behavior of the EPR peak. Figure~7 shows the magnetic resonance spectra obtained at different temperatures for the film (CoFeB)$_{25}$(Al$_2$O$_3$)$_{75}$ in magnetic field applied in the film plane. As~can be seen from the figure, with temperature decreasing from 296 to 60~K, the intensity of the EPR peak increases significantly. A similar, but less pronounced increase in the intensity is observed for the FMR peak (``superparamagnetic'' resonance peak in other terms). In this regard, the evolution of the spectra in the temperature range $60{-}296$~K looks quite natural.

\begin{figure}
\centering
\includegraphics[width=0.9\columnwidth]{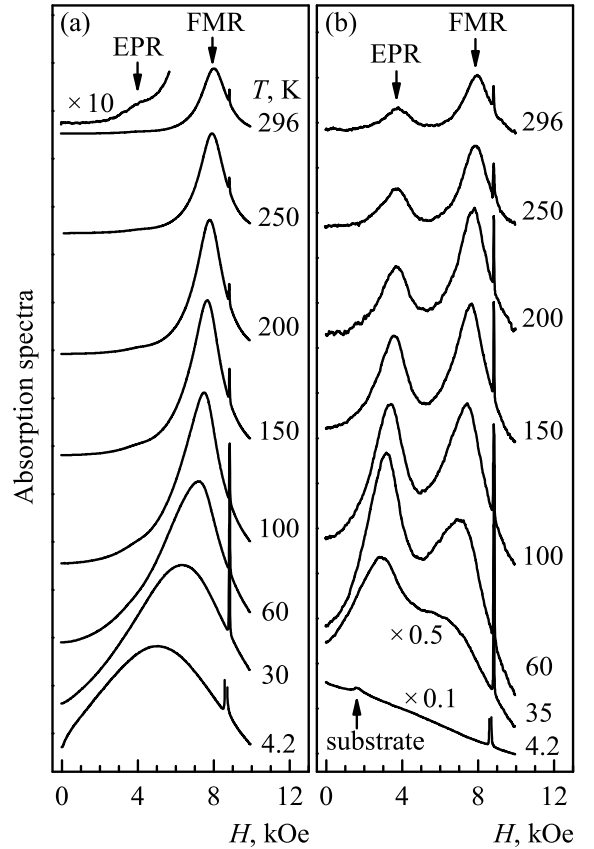}
\caption{Spectra of the film (CoFeB)$_{25}$(Al$_2$O$_3$)$_{75}$ at different temperatures $T=4.2{-}296$~K in the case of in-plane field at frequency $f\approx25$~GHz for (a)~transverse and (b)~longitudinal geometries of resonance excitation. In the case of longitudinal geometry, the vertical scale of the spectra is enlarged 20 times with respect to the transverse geometry.}
\end{figure}

However, with a further decrease of temperature from 60 to 4.2~K, the form of the spectra changes significantly. In the case of transverse resonance excitation geometry ($\mathbf{h}\perp\mathbf{H}$), it is seen that the FMR peak strongly broadens and shifts to weak fields. The EPR peak observed in the longitudinal geometry ($\mathbf{h}\parallel\mathbf{H}$) vanishes. At the same time, a strong absorption appears in the vicinity of $H =0$. A small peak, manifested in weak fields at $T=4.2$~K, is associated with PM impurities in the substrate \cite{Drov2022-2}.

The observed transformation of the spectra in the low temperature region is probably explained by a decrease of thermal fluctuations in the system of PM ions and FM nanogranules and the formation of larger magnetically ordered clusters coupled by exchange and magnetodipole interactions. This is accompanied by suppression of EPR of individual ions and the formation of collective oscillation modes with a wide frequency spectrum due to the strong inhomogeneity of the system and fluctuations of local anisotropy. This scenario explains the observed disappearance of the EPR peak and the strong broadening of the FMR line in the low temperature region.

Further, it can be assumed that with an increase of the FM phase content in the nanocomposite, the formation of macroscopic magnetically ordered clusters begins at higher temperatures. This explains the expansion of the temperature range with ``anomalous'' behavior of the EPR peak, as well as the disappearance of this peak in samples with the FM phase content above the percolation threshold.

It can also be expected that due to the high degree of disorder, the low-temperature state of the system has the features of spin (cluster) glass. This state is characterized by a high density of local energy minima corresponding to various magnetic configurations of the system. The density of these minima (quasi-equilibrium states) decreases with increasing magnetic field when the system approaches the saturation. Thus, the absorption in the vicinity of $H=0$ observed in low-temperature spectra at $\mathbf{h}\parallel\mathbf{H}$ can be associated with the excitation of transitions between various quasi-equilibrium states of the system.

\section{Conclusion}

Nanogranular metal-insulator films M$_x$D$_{100-x}$ with different compositions (M = Fe, Co, CoFeB; D = Al$_2$O$_3$, SiO$_2$, LiNbO$_3$) and various FM metal contents $x$ were studied by magnetic resonance. The experimental spectra of the films contain the FMR line from the ensemble of FM granules, as well as an additional absorption peak with effective $g$-factor $g\approx4.3$, which we associate with resonance at PM centers present in the insulating space between FM granules. Fe$^{3+}$ and Co$^{2+}$ ions dispersed in the insulating matrix during the deposition of the films can serve as such centers.

With an increase of the FM phase content, the observed EPR line ($g\approx4.3$) demonstrates an additional shift depending on the orientation of the magnetic field with respect to the film plane. The correlation of this shift with the demagnetization field of the film $4\pi M$ has been experimentally established. When magnetic field is applied in the film plane, the EPR line is shifted to weaker fields by an amount of $\approx 4\pi M/3$ relative to its position in the limit of low FM phase contents. On~the contrary, for the normal orientation of the field, the EPR peak is shifted to stronger fields by $\approx 8\pi M/3$. This behavior can be explained by magnetodipole fields acting on PM centers from the ensemble of FM granules: the demagnetizing field $-4\pi M$, which arises in a normally magnetized film, and the Lorentz field $4\pi M/3$, which is independent of the external field orientation.

Due to fluctuations of the local dipole and exchange fields at PM centers, the EPR peak is manifested not only in the usual transverse, but also in the longitudinal geometry of the resonance excitation.

The presence of magnetic interactions in the system of PM ions and FM granules also leads to a peculiar temperature dependence of the EPR peak amplitude. When cooling from the high temperature region, the intensity of the EPR line first rises due to an increase in the susceptibility of PM ions. However, at low temperatures, the weakening of thermal fluctuations leads to the formation of macroscopic coupled clusters in the system of PM ions and FM granules. In this situation, the EPR peak from individual ions decreases until its complete disappearance, when the magnetically ordered state extends to the entire film.

Thus, in this paper we have shown that PM ions dispersed in an insulating matrix in metal-insulator nanogranular composites can serve as markers of magnetic interactions present in the system. These interactions are manifested while studying the behavior of the EPR line of the dispersed ions.

\section*{Funding}

The work was carried out within the framework of a~state assignment and was financially supported by the Russian Science Foundation (project no. 22-29-00392).

\section*{Conflict of interest}

The authors declare that they have no conflicts of interest.

\end{document}